# Double-slit interference with charged particles. Density matrices and decoherence from time-dependent quantum Monte Carlo.

Ivan P. Christov

Physics Department, Sofia University, 1164 Sofia, Bulgaria

Email: ivan.christov@phys.uni-sofia.bg

**Abstract**

In this paper we apply the time-dependent quantum Monte Carlo (TDQMC) method to explore a modified single- and double-slit diffraction of matter waves. By using a simplified model of two electrons prepared in the ground state of an atom (molecule) and then suddenly released we are able to calculate the diffraction patterns in one spatial dimension in close correspondence with the numerically exact results. Through the Coulomb repulsion the one electron serves as an environment for the other thus introducing decoherence in the quantum state which is easily quantified. It is demonstrated that the set of single particle wave functions yield by TDQMC can be used to directly construct density matrix for that particle without tracing out the other constituents from the density matrix of the whole system. In this way it is possible to build explicitly time-dependent density matrices for different components of a complex quantum system straight within the TDQMC algorithm which may widen our understanding of quantum dynamics in many-body systems.

## 1. Introduction

The wave-particle duality of microscopic objects is a key ingredient of quantum physics. The wave-like properties of matter formulated first by L. de Broglie have established a vast area of studies where the phase of the matter waves is of importance. Recently diffraction and interference effects have been observed experimentally with typical quantum objects like electrons and neutrons [1,2] as well as with mesoscopic objects such as $C_{60}$ [3] and $C_{70}$ [4]. The double-slit type of experiments which are among the simplest and most general quantum experiments have demonstrated undisputedly the significance of the information about the quantum states of a microscopic object which can be extracted from the interference patterns. One important piece of information which is attainable from such experiments concerns the degree of coupling of the quantum particle with its environment, also known as decoherence, which may partially or completely destroy the interference [5]. The effect of decoherence on matter waves has been studied also as a basic mechanism for the quantum-classical transition [6]. Theoretically the effects of decoherence are treated by coupling the quantum system under consideration to a model environment, e.g. a set of harmonic oscillators, which leads to master equation for the reduced density matrix [7] where the environment degrees are not considered explicitly. Typically that coupling leads to an exponential damping of the off-diagonal elements of the density matrix [8] which manifests itself as a quenching of the interference fringes in the double-slit setup.

Recently another experiment has shown that two protons produced by double photoionization of hydrogen molecule may form a charged particle/slit system where each photo-ionized electron constitutes a minimum environment for the observation of decoherence effects related to the other electron [9]. In that experiment the interference fringes observed in the angular distribution of one electron are lost due to its Coulomb interaction with the other electron. Here we employ the recently devised time dependent quantum Monte-Carlo method [10-12] to calculate the ground state and the time evolution of the wave functions in two-electron atom (molecule) after the electrons have been released suddenly and they interact with each other through a Coulomb potential in one spatial dimension. In other words, in order to simplify the charged-particle single-slit (double-slit) experiment we neglect the interaction of the atom (molecule) with the powerful external optical pulse which in fact causes the release of the electrons. Since the protons are much heavier than the electrons we assume that they are at rest at some distance from each other, thus constituting the two slits through the corresponding ground states of the electrons. In coordinate representation, after the ground state wave functions of the two electrons in the atom (molecule) are prepared, the Coulomb field that is due to the protons is switched off, which causes free diffraction of the electron wave packets where the only interaction left between them is due to their mutual Coulomb interaction.

The time-dependent quantum Monte Carlo (TDQMC) method employs configurations of particles and guiding waves to describe the ground state and the time evolution of many-body quantum systems in physical space. Each physical particle is represented in TDQMC by a statistical ensemble of walkers, which reproduces the quantum probability distribution of that particle. Each walker is guided by a separate de Broglie-Bohm guide wave which obeys its own time-dependent Schrödinger equations (TDSE) in physical space where the particle interactions are accounted for using explicit non-local potentials which involve positions of walkers which represent the rest of the particles. In this way TDQMC treats symmetrically the Monte-Carlo walkers and the guide waves where besides the walker's guidance by the waves there is also a reverse action of the walker's trajectories exhibited by the waves through the potentials in the Schrödinger equations. This distinguishes TDQMC from other quantum Monte Carlo methods where the stochastic component in the walker's motion dominates while the guide function needs to be guessed in advance. Also, it is essential that since in TDQMC both walkers and guide waves evolve in physical space there are no difficulties due to wave functions defined in multi-dimensional configuration space. In fact a very good example of how to eliminate the configuration space in quantum mechanics is provided by the well known Hartree-Fock approximation which has historically been the first instrumental approach to replace the many-body Schrödinger equation by a set of single particle Schrödinger equations in physical space. The time-dependent Hartree-Fock method appears as a limiting case of TDQMC where there is no correlation between the motion of the individual particles.

During the preparation of the initial state each Monte-Carlo walker experiences quantum drift and diffusion by sampling its own probability density distribution given by the modulus squared of the corresponding guide wave. Concurrently a set of coupled time-dependent Schrödinger equations for the guide waves is solved in imaginary time. In some cases it is advantageous to apply an appropriate importance sampling technique in order to efficiently direct the walkers towards parts of space with higher probability density. Events of birth/death of walkers and guide waves (also called branching) can be used to significantly reduce the relaxation time and to improve the accuracy of the final distribution. Once the ground state is established, real-time quantum dynamics can be studied which involves real time solution of the set of time-dependent Schrödinger equations together with the first order differential equations for the walker's motion. For recent reviews on applying quantum Monte Carlo methods in different contexts see Refs. [13,14].

## 2. Methods

In time-dependent quantum Monte Carlo the guide waves $\varphi_i^k(\mathbf{r}_i,t)$ obey a set of coupled TDSE for the $k$-th pair of walker/guide wave from the $i$-th electron ensemble [11,12]:

$$i\hbar\frac{\partial}{\partial t}\varphi_i^k(\mathbf{r}_i,t)=\left[-\frac{\hbar^2}{2m}\nabla_i^2+V_{e-n}(\mathbf{r}_i)+\sum_{j\neq i}^{N}V_{e-e}^{eff}[\mathbf{r}_i-\mathbf{r}_j^k(t)]\right]\varphi_i^k(\mathbf{r}_i,t), \quad (1)$$

where $V_{e-n}(\mathbf{r}_i)$ is the electron-nuclear potential and $V_{e-e}^{eff}[\mathbf{r}_i-\mathbf{r}_j^k(t)]$ is the effective electron-electron potential experienced by the walkers from the $i$-th electron ensemble due to the walker's trajectories belonging to the rest of the electrons $\mathbf{r}_j^k(t)$. In this way the many-body Hamiltonian is reduced to one-body Hamiltonians where the degrees of freedom of the rest of the particles are expressed through their trajectories. The effective nonlocal electron-electron potential in Eq.(1) can be represented as a Monte Carlo sum over the Coulomb potentials $V_{e-e}[\mathbf{r}_i-\mathbf{r}_j^l(t)]$ and a nonlocal kernel function $K\left[\mathbf{r}_j^l(t)-\mathbf{r}_j^k(t)\right]$:

$$V_{e-e}^{eff}[\mathbf{r}_i-\mathbf{r}_j^k(t)]=\frac{1}{Z_j^k}\sum_{l=1}^{M}V_{e-e}[\mathbf{r}_i-\mathbf{r}_j^l(t)]\;K\left(\frac{\left|\mathbf{r}_j^l(t)-\mathbf{r}_j^k(t)\right|}{\sigma_j^k\left(\mathbf{r}_j^k,t\right)}\right), \tag{2}$$

where:

$$Z_j^k=\sum_{l=1}^{M}\;K\left(\frac{\left|\mathbf{r}_j^l(t)-\mathbf{r}_j^k(t)\right|}{\sigma_j^k\left(\mathbf{r}_j^k,t\right)}\right), \tag{3}$$

is the weighting factor. The width $\sigma_j^k\left(\mathbf{r}_j^k,t\right)$ of the nonlocal kernel in Eq. (2) is a measure for the characteristic length of nonlocal quantum correlations for the *j*-th electron. The Monte Carlo trajectories evolve according to the de Broglie-Bohm guiding equation:

$$\mathbf{v}(\mathbf{r}_i^k)=\frac{\hbar}{m}\mathrm{Im}\left[\frac{1}{\Psi^k(\mathbf{r}_1,...,\mathbf{r}_N,t)}\nabla_i\Psi^k(\mathbf{r}_1,...,\mathbf{r}_N,t)\right]_{\mathbf{r}_j=\mathbf{r}_j^k(t)}, \tag{4}$$

where i=1,…,N; k=1,…,M denote the electrons and the walkers for each electron, respectively. For no spin variables in the Schrödinger equation the many-body wave function can be represented as an anti-symmetrized product (Slater determinant or a sum of Slater determinants) of the individual guide waves:

$$\Psi^k(\mathbf{r}_1,\mathbf{r}_2,...,\mathbf{r}_N,t)=A\prod_{i=1}^{N}\varphi_i^k(\mathbf{r}_i,t)\,. \tag{5}$$

However, since we consider the propagation of both walkers and guide waves in physical space we shall ignore for a moment the symmetrization (anti-symmetrization) in Eq.(5) and assume a simple product state, which implies that each walker is guided by its own wave:

$$\mathbf{v}(\mathbf{r}_i^k) = \frac{\hbar}{m} \mathrm{Im}\left[ \frac{1}{\varphi_i^k(\mathbf{r}_i, t)} \nabla_i \varphi_i^k(\mathbf{r}_i, t) \right]_{\mathbf{r}_i = \mathbf{r}_i^k(t)} . \quad (6)$$

The many-body probability distribution in configuration space can then be calculated (if needed) as described in [15].

In practice, the parameters $\sigma_j^k\left(\mathbf{r}_j^k, t\right)$ in Eq.(2) and Eq.(3) can be determined by variationally minimizing the ground state energy of the quantum system. Note that here $\sigma_j^k\left(\mathbf{r}_j^k, t\right)$ characterizes the spatial quantum nonlocality unlike in other cases where nonlocal causality is of concern (e.g. in Einstein-Podolsky-Rosen pairs [16]). It is assumed here that the nonlocal quantum correlation length is simply proportional (to the first approximation) to the kernel density estimation (KDE) bandwidth of the Monte Carlo ensemble for each electron $\sigma_j^k\left(\mathbf{r}_j^k, t\right) = \alpha_j . \sigma_{j,KDE}^k\left(\mathbf{r}_j^k, t\right)$ [12]. Depending on the technique used for KDE the bandwidth $\sigma_{j,KDE}^k\left(\mathbf{r}_j^k, t\right)$ may be simply proportional to the standard deviation of the Monte Carlo sample (Silverman's "rule of thumb" KDE). Since in Silverman's KDE $\sigma_{j,KDE}^k\left(\mathbf{r}_j^k, t\right)$ contains a weak dependence on the number of Monte Carlo walkers it is more convenient and physically justified to accept that the nonlocal quantum correlation length is proportional (to the first approximation) to the standard deviation $\sigma_j$ of the Monte Carlo sample for the *j*-th electron:

$$\sigma_j^k\left(\mathbf{r}_j^k, t\right) = \alpha_j . \sigma_j\left(t\right) . \quad (7)$$

Note that depending on the symmetry of the sample both the (KDE) bandwidth and the standard deviation may vary in different directions and therefore the use of covariance matrix in Eq.(7) might be preferable.

It is important to point out that the set of TDQMC equations (Eq.(1) and Eq.(2)) is essentially nonlinear and the only linear limiting case is for $\alpha_j \rightarrow 0$ where the nonlocal quantum correlation length tens to zero and each walker from given electron ensemble interacts with only one walker from the ensembles representing the rest of the electrons [10]. It is noteworthy that the opposite case where $\alpha_j \to \infty$ corresponds to the mean field (Hartree-Fock) approximation where the equations for the guide waves are nonlinear because the Monte Carlo sum in Eq.(2) contains an implicit dependence on the probability densities.

Since the walker’s distributions in space are somewhat privileged in that they reproduce the quantum probability densities, it seems from the guiding equation (Eq.(6)) that due to the locality of the gradient operator much of the information contained in the wave functions $\varphi_i^k(\mathbf{r}_i,t)$ remains unused and it is actually lost at each next step of time propagation. However, here we show that the variety of guide waves and their time evolution can be used very efficiently to construct approximate density matrices in coordinate representation for the different electrons and, in general, for any quantum particle under consideration. First let us mention that the effective electron-electron potential in Eq.(1) may be very different for the different guide waves and after the ground state preparation we would end up with ensembles of waves with certain statistical distribution for each physical particle. Therefore for the interacting electron the ground state may be considered as a mixed state where we know explicitly the waves which participate in the mixture together with their classical probabilities which follow the probability distribution of the walkers in space. Thus, an ensemble average over the guide waves of the *i*-th electron would read:

$$\left\langle \hat{A}_i\left(t\right)\right\rangle = \frac{1}{M}\sum_{k=1}^{M}\int \varphi_i^{k*}(\mathbf{r}_i,t)\hat{A}_i\left(\mathbf{r}_i,\mathbf{r}_i',t\right)\varphi_i^k(\mathbf{r}_i',t)d\mathbf{r}_i d\mathbf{r}_i' = Tr\left(\rho_i A_i\right) , \qquad (8)$$

where the density matrix is defined by the guide waves:

$$\rho_i\left(\mathbf{r}_i,\mathbf{r}_i',t\right) = \frac{1}{M}\sum_{k=1}^{M}\varphi_i^{k*}(\mathbf{r}_i,t)\varphi_i^k(\mathbf{r}_i',t) \quad . \qquad (9)$$

Since during the ground state preparation all guide waves are kept normalized we have from Eq.(9) for the normalization of the density matrix:

$$Tr\left[\rho_i\left(\mathbf{r}_i,\mathbf{r}_i',t\right)\right]=\int \varphi_i^{k*}(\mathbf{r}_i',t)\rho_i\left(\mathbf{r}_i,\mathbf{r}_i',t\right)\varphi_i^k(\mathbf{r}_i,t)d\mathbf{r}_i d\mathbf{r}_i' = 1 \tag{10}$$

The key result here is that the TDQMC methodology basically allows the construction of density matrices for a subsystem of larger quantum system directly using the set of wave functions provided by the solution of the coupled Schrödinger equations (Eq.(1)) instead of solving the master equation and/or partially tracing the density matrix of a larger system. The analysis of the time evolution of on- and off-diagonal elements of the density matrix provides detailed information on e.g. the processes of decoherence for the various degrees of freedom, time dependent correlations, etc. Note that the diagonal elements of the density matrix Eq.(9) have already been used previously to accurately calculate quantum averages such as the dipole moment of an atom exposed to intense electromagnetic field [17].

## 3. Results

In order to estimate the accuracy of the TDQMC predictions in the examples below we employ a model of a two-electron atom and a two-electron molecule in one spatial dimension. This simplified model allows us to compare the TDQMC results with the numerically exact solution of the two-body Schrödinger equation with soft core potentials which has been widely used in the literature (e.g. [18]) (atomic units are used henceforth):

$$i\frac{\partial}{\partial t}\Psi(x_1,x_2,t)=\left[H_0+V_{e-e}\left(x_1-x_2\right)\right]\Psi(x_1,x_2,t), \tag{11}$$

where the free Hamiltonian operator in coordinate representation reads:

$$H_0=-\frac{1}{2}\frac{\partial^2}{\partial x_1^2}-\frac{1}{2}\frac{\partial^2}{\partial x_2^2}-\frac{a}{\sqrt{1+x_1^2}}-\frac{a}{\sqrt{1+x_2^2}} \tag{12}$$

and the electron-electron interaction potential is given by:

$$V_{e-e}\left(x_1-x_2\right)=\frac{b}{\sqrt{1+(x_1-x_2)^2}} \quad . \tag{13}$$

In Eq.(12) and Eq.(13) the parameters $a$ and $b$ determine the strength of the electron-nuclear and electron-electron interaction, respectively. It is assumed that $a$=2 during the preparation of the ground state and $a$=0 for the real time diffraction of the resultant electron wave packets. From the exact wave function $\Psi(x_1,x_2,t)$, the exact trajectories $x_i^e\left(t\right)$ can be calculated by numerically integrating the equations:

$$\frac{dx_i^e}{dt}=\mathrm{Im}\left[\frac{1}{\Psi(x_1,x_2,t)}\nabla_i\Psi(x_1,x_2,t)\right]_{x_1=x_1^e(t);x_2=x_2^e(t)} \quad . \tag{14}$$

The corresponding set of TDQMC equations (Eq.(1), Eq.(2)) are:

$$i\hbar\frac{\partial}{\partial t}\varphi_i^k(x_i,t)=\left[-\frac{1}{2}\nabla_i^2-\frac{a}{\sqrt{1+x_i^2}}+\sum_{j\neq i}^{N}V_{e-e}^{eff}[x_i-x_j^k(t)]\right]\varphi_i^k(x_i,t)\text{, } i=1,2;\ k=1,M, \tag{15}$$

where the effective electron-electron potential reads:

$$V_{e-e}^{eff}[x_i-x_j^k(t)]=\frac{1}{Z_j^k}\sum_{l=1}^{M}\frac{b}{\sqrt{1+\left[x_i-x_j^l(t)\right]^2}}\exp\left(-\frac{\left|x_j^l(t)-x_j^k(t)\right|^2}{2\sigma_j^k\left(x_j^k,t\right)^2}\right) \tag{16}$$

Before we discuss the results from the concrete calculations one needs to point out that since the quantum Monte Carlo methods are of statistical nature, the larger number of walkers (and guide waves for TDQMC)) ensures better accuracy of the final result. Unlike in the standard quantum Monte Carlo (e.g. its diffusion version, Ref.[13]), there is no need to start with accurate guide function in TDQMC. Instead, the imaginary-time evolution of almost arbitrary reasonable initial set of guide functions adjusts those to their optimal shape, together with the optimal positions of the corresponding walkers. Here we begin with a single-slit charged-particle diffraction where two electrons with opposite spin are first prepared in a symmetric ground state (s-state) as described previously, and are then suddenly released for free diffraction. We consider full scale Coulomb repulsion between the electrons where $b=1$ in Eq.(13). It was found that the minimum ground state energy of the model atom is reached for $\alpha_1 = \alpha_2 = 0.6$ in Eq.(7). Figure 1(a) depicts four randomly chosen probability densities $\left|\varphi_i^k(x_i,t)\right|^2$ for one of the electrons after the ground state has been established. Figure 1(b) shows these densities after diffraction. It is seen that these densities differ significantly due to the different nonlocal effective potentials experienced by the waves in the TDQMC equations (Eq.(15) and Eq.(16)).. It should be noted that all densities would become identical if we set $\alpha_1, \alpha_2 \to \infty$ in Eq.(7) and Eq.(16) which reinstates the mean field (Hartree-Fock) approximation where all waves see the same electron-electron potential given by an weighted average in Eq.(16) and all walkers for a given electron are guided by the same guide wave. As mentioned earlier, in the latter case the nonlinearity is most pronounced in the TDQMC equations, Eq.(15), which is opposed to the linear "ultra-correlated" case where $\alpha_1, \alpha_2 \to 0$ and the nonlocal quantum correlation effects are neglected. Figures 1(c),(d) present the case of a molecule where the two protons are separated by 8 a.u. It is important to stress that the values of the parameters $\alpha_j$ in Eq. (7) are kept constant during the whole time evolution of the many-body system, so that the non-local quantum correlation length is determined exclusively by the quantum diffusion in presence of interactions between the particles which reflect on the standard deviation of the walkers' ensemble for each electron.

Figure 2(a),(b) shows the initial and the final one-electron probability densities from the TDQMC and the exact calculations for single-slit diffraction of interacting electrons where the TDQMC curve has been calculated either by performing KDE over the walker distribution or by using the diagonal elements of the density matrix $\rho(x,x,t)$ in Eq.(9). The green line represents the KDE curve calculated using an accurate KDE algorithm [19] while the red line shows the density matrix result. It is seen that the two

curves fit quite well with the exact result (blue line), which manifests the good accuracy achievable by the TDQMC method for the diffraction of charged particles. Of course, the overlap of the exact curves with the TDQMC predictions from the density matrix is better because the guide waves basically offer much richer statistics than the point-like walkers. As expected, due to the Coulomb repulsion and the strong overlap of the ground state electron wave functions (Fig.1(a)) rapid separation of the two electrons in space occurs. In order to estimate the accuracy of our calculation as function of time we show in Fig.2(c) with red line the deviation between the TDQMC probability density and the exact result as compared with that deviation for the linear ultra-correlated $(\alpha_1, \alpha_2 \to 0)$ case (blue line). As an appropriate measure for the deviation of the probability density functions in time here we employ the modulus of the difference between the approximate and the exact curves which is next integrated along the x axis. In this way, the resulting deviation quantifies the error in the TDQMC calculations. It is seen from Fig.2(c) that the TDQMC predictions remain closer to the exact result for the whole integration interval while the deviation of the ultra-correlated case (where there is stronger repulsion between the walkers) increases with time.

Next, we present the results for the double-slit diffraction where the two protons are separated by a constant distance of 8a.u. The initial and the final one-electron probability densities from TDQMC are compared with the exact results in Fig.3, with the same colors as in Fig.2, for different strengths of the Coulomb repulsion: *b*=0.02 in Fig.3(a),(b), *b*=0.1 in Fig.3(d),(e), and *b*=1 in Fig.3(g),(h). These strengths are chosen such that the visibility of the interference fringes vary from good in Fig.3(b) (where the decoherence effects experienced due to the other electron are not significant) to poor in Fig.3(e) (for moderate decoherence), and to totally missing in Fig.3(h) where the two peaks are due to Coulomb repulsion rather than interference (cf. Fig.2(b)). It is seen that there is again a good overlap between the exact curves and the TDQMC predictions, especially with those obtained from the density matrix. Since the interference is very sensitive to any disturbances of the amplitudes and the phases of the waves, the results presented here indicate that despite the huge number of walkers and guide waves which participate in the calculation (up to 216 000 in this work) the final probabilities show good agreement with the solution of the two-body Schrodinger equation. In fact, the comparison between the time dependent deviations drawn in Fig.3(c),(f),(i) tells us that the absolute error of the TDQMC solution is an order of magnitude smaller for the double-slit diffraction (see Fig.2(c) which can be attributed to the fact that for the double-slit the particles are further away from each other for most of the time and they experience weaker electrostatic field.

Figure 4(a)-(d) present the module of the density matrices in the coordinate representation for single-slit diffraction where *b*=1. It is seen that the initial state is characterized by a symmetric pattern which is then transferred to a two lobe distribution mainly due to the Coulomb repulsion. In fact, the diagonals of the patterns in Fig.4(a),(b) correspond to the red lines in Fig.2(a),(b). It is essential that the evolution of the density matrix during the diffraction represents the most complete description of the quantum state of the single electron. For example, the off-diagonal elements of the density matrix are sensitive to the phases of the different guide waves and therefore these characterize the loss of coherence due to the Coulomb interaction between the electrons. One can get an idea on the effects of decoherence by simply calculating the average of the moduli of the anti-diagonal elements of the density matrix for different moments of time, as shown in Fig.5 for an electron released from a two-electron atom at t=0. The rapid drop of the blue curve evidences that the degree of mutual coherence within the ensemble of guide waves for one electron decreases as the diffraction progresses. It is seen that here the coherence drops by more than 50% for about $10^{-16}$ seconds. This is in a reasonable correspondence with the numerically exact result (black line) obtained from the direct integration of the two-body time-dependent Schrodinger equation (Eq. (11) and next using the standard definition of the reduced (single-electron) density matrix expressed through the resulted wave-functions:

$$\rho\left(x,x',t\right)=\frac{\int\Psi(x,x_1,t)\Psi^*(x',x_1,t)dx_1}{\iint\Psi(x,x_1,t)\Psi^*(x,x_1,t)dx_1dx}, \qquad (17)$$

where in the denominator stays the trace of density matrix for normalization. For the ultra-correlated case (green line) the drop of the coherence is even faster as expected.

This opposes the case of Hartree-Fock approximation where there is only one guide wave for each electron and high coherence is maintained in due course (red line in Fig.5).

## 4. Conclusions

Here, the recently devised time-dependent quantum Monte Carlo method is used to analyze single- and double-slit diffraction of electron wave packets prepared in advance as ground states of two-electron atom (molecule). The main goal is to study the effects of decoherence as the two electrons are released at time zero where each electron serves as an environment for the other one, and to compare the results with the numerically exact results obtained from the direct solution of the two-body Schrodinger equation. It is shown that for optimally chosen length of nonlocal quantum correlations the TDQMC result is very close to the exact result, which is next used to introduce a density matrix in coordinate representation based on the set of waves obtained from the TDQMC calculation. It is demonstrated that the density matrix may describe a single quantum particle without tracing out subsystems from the density matrix of the whole system. In this way the explicit construction of density matrices for different components of a complex quantum system directly within the TDQMC algorithm may provide a valuable tool for exploration of quantum effects in many-body systems that have been beyond reach until now.

Since TDQMC uses some concepts related to the de Broglie-Bohm theory it would be reasonable to clarify whether the TDQMC guide waves are related to the so called “conditional wave functions” used in Bohmian mechanics [20]. Despite the seeming similarities between the two the answer would rather be “not necessarily”. The “conditional wave function” is primarily aimed at describing a subsystem of a larger quantum system in a reduced dimensionality space where the quantum trajectories due to the conditional wave should follow precisely the exact ones derived from the many-body dynamics. Since in TDQMC the Monte Carlo walkers do not represent real physical particles (unlike in Bohmian mechanics) the important outcome from the calculation is the walker’s spatial distribution which yields the one-body (or many-body) probability density in terms of standard quantum mechanics. Hence, the TDQMC trajectories are not expected to reproduce exactly those obtained from the many-body Schrodinger equation which can be calculated for the simplest cases only. In fact, it can be shown that the TDQMC trajectories may increasingly deviate from the exact ones while at the same time the walker distribution stays in close agreement with the exact result. In general, the derivation of exact Bohmian trajectories from a reduced dimensionality calculation seems feasible for very weak interactions (entanglement) only where any departure due to combined action of nodes of the many-body state, quantum chaos, and nonlinearities is negligible. In a realistic calculation even small fluctuations which may occur due to e.g. numerical error may result in a significant deviation of the quantum trajectories from their presumed positions.

**Acknowledgments**

The author gratefully acknowledges support from the Bulgarian National Science Fund under grant FNI T02/10.

**Figure captions:**

**Figure 1**. Moduli square of four randomly chosen guide waves for one electron in a two-electron atom at ground state (a), and after sudden release and diffraction (b). The same for two-electron molecule (c) and (d), respectively.

**Figure 2.** Probability density for a single electron of two-electron atom at ground state (a), and after diffraction due to sudden release (b). Blue lines-exact result, red lines–from TDQMC density matrix, green lines-from kernel density estimation over walker's distribution in space. (c)-deviation from the exact probability density for the optimized nonlocal correlation length (red) and for the ultracorrelated case (blue).

**Figure 3**. Probability densities of a single electron at ground state of two-electron molecule for different coupling between the electrons due to Coulomb repulsion: *b*=0.02-(a); *b*=0.1-(d), and *b*=1-(g), and after diffraction due to sudden release (b), (e), (h), respectively. (c),(f),(i)-deviation from the exact probability densities for the optimized nonlocal correlation length (red) and for the ultracorrelated case (blue).

**Figure 4.** Density matrix in the coordinate representation for single electron in a two-electron atom at ground state (a) and after diffraction (b).

**Figure 5.** Degree of coherence for an interacting electron released from a two-electron atom as function of time during diffraction. The lines represent the time evolution of the anti-diagonal sum of the one-electron density matrix. Blue line –TDQMC result (Eq. (9); black line –exact result (Eq. (17)); green line –ultra-correlated result ($\alpha_1, \alpha_2 \to 0$); red line –Hartree result ($\alpha_1, \alpha_2 \to \infty$).

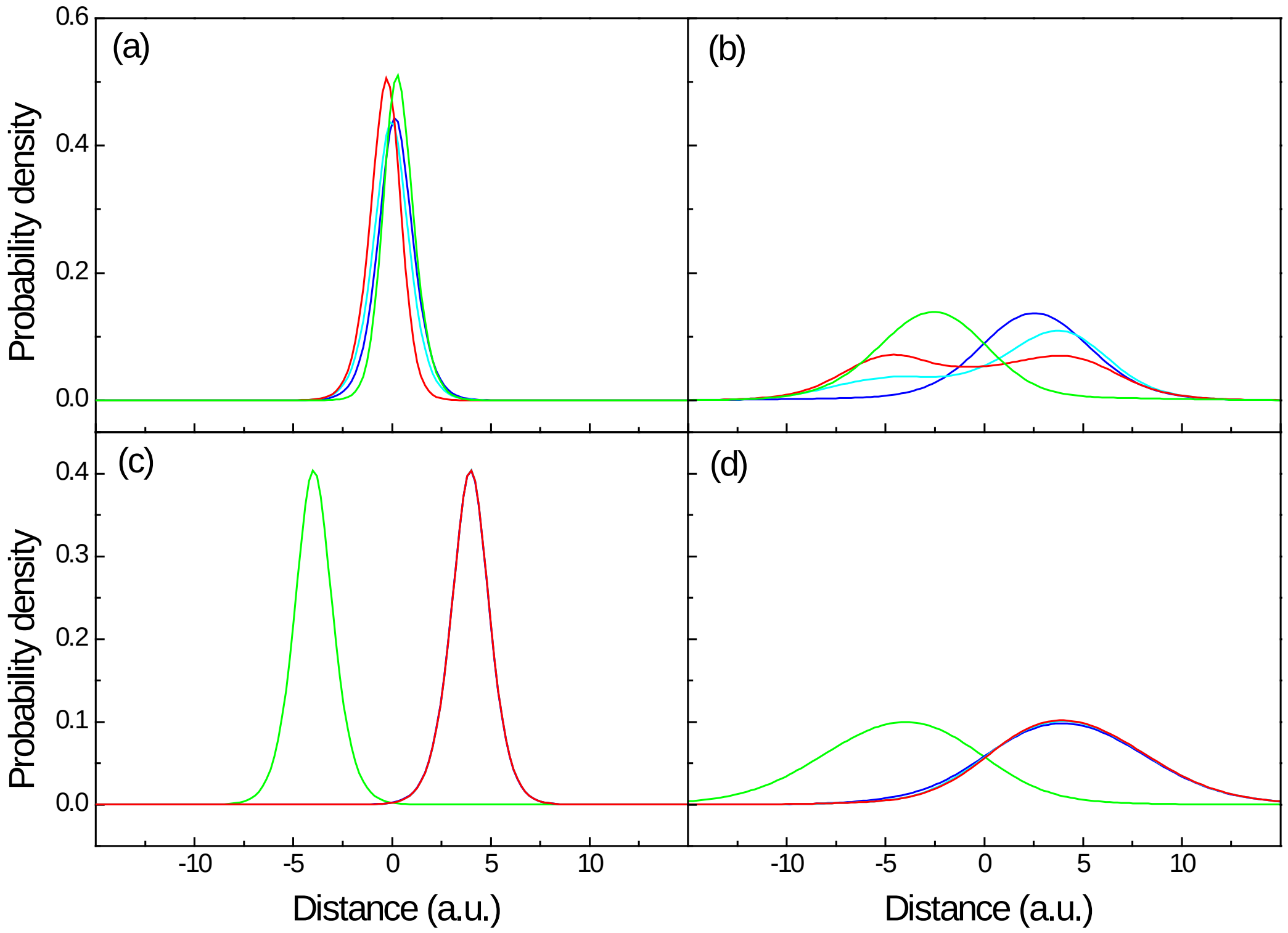

Christov, Figure 1

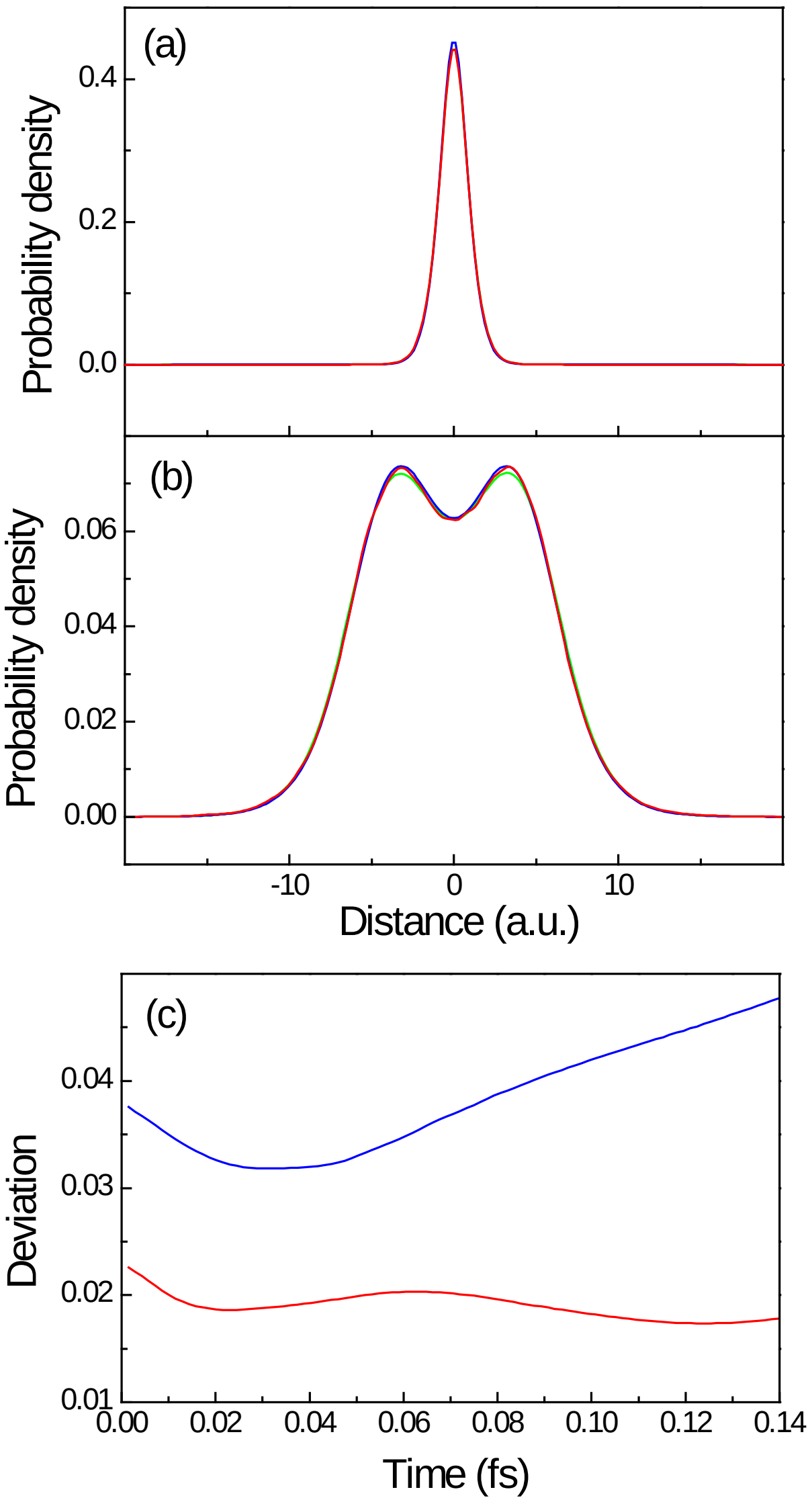

Christov, Figure 2

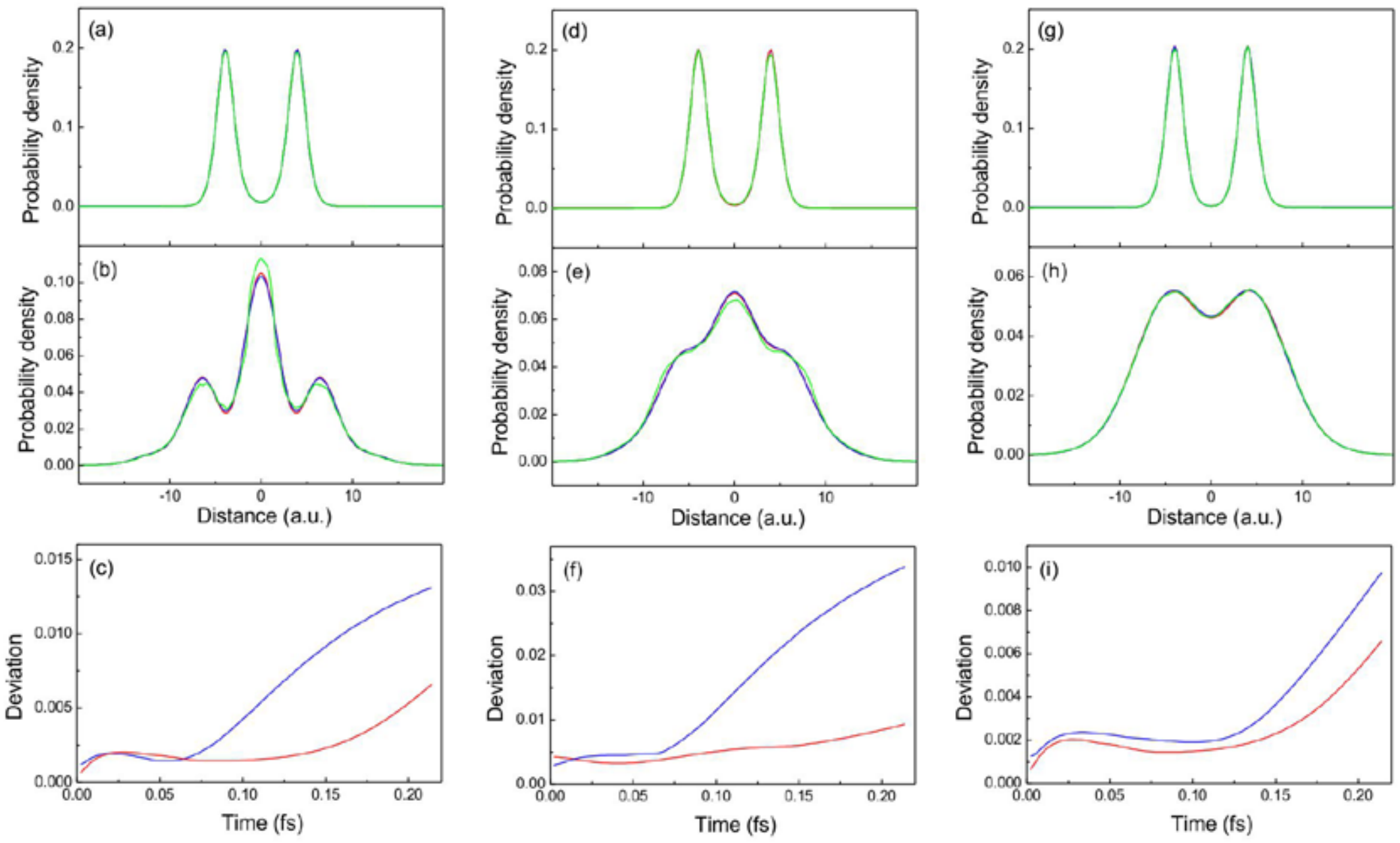

Christov, Figure 3

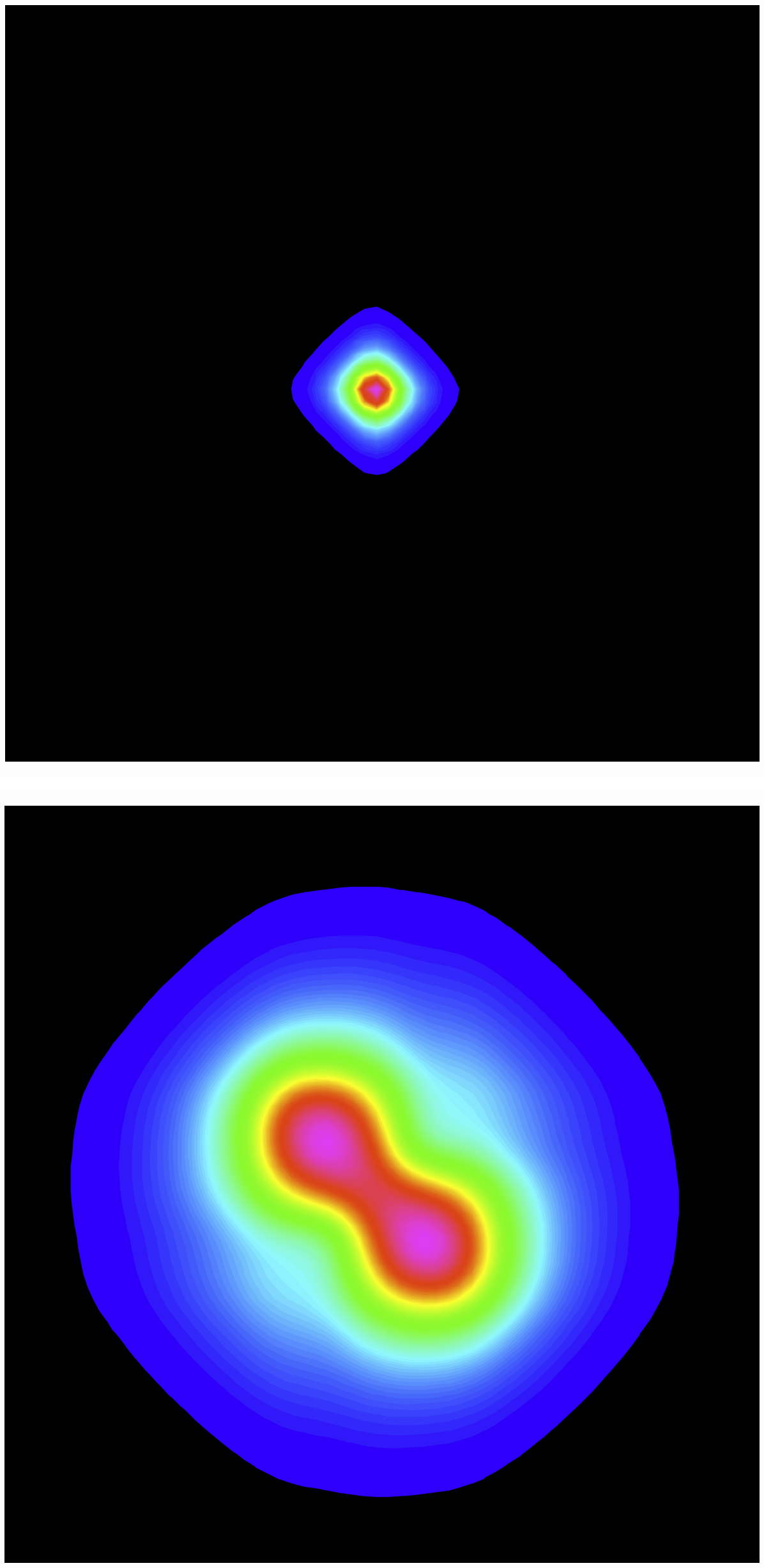

(a)

(b)

Christov, Figure 4

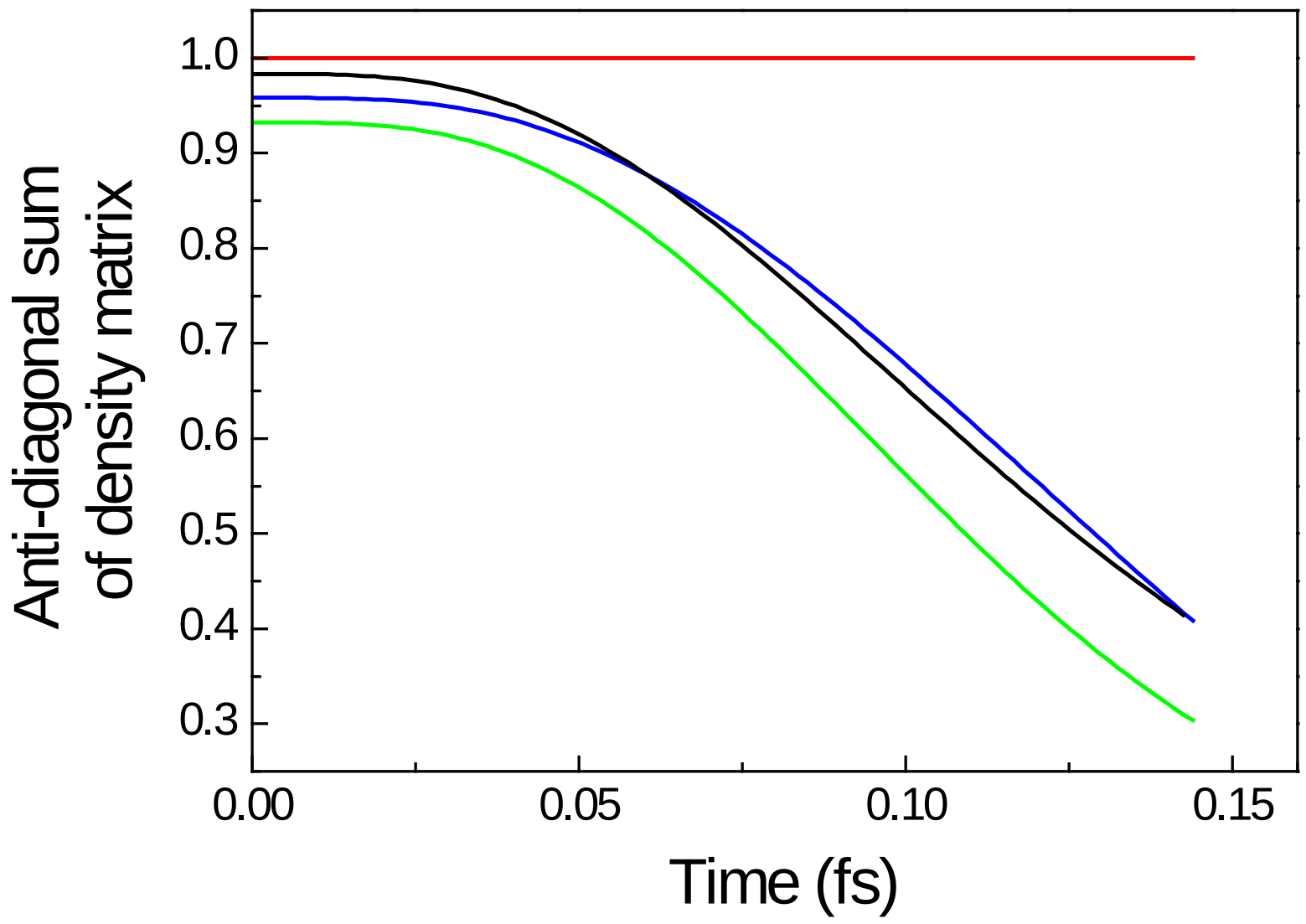

Christov, Figure 5